\input harvmac
\input graphicx
\input color

\def\Title#1#2{\rightline{#1}\ifx\answ\bigans\nopagenumbers\pageno0\vskip1in
\else\pageno1\vskip.8in\fi \centerline{\titlefont #2}\vskip .5in}

%
%
\ifx\includegraphics\UnDeFiNeD\message{(NO graphicx.tex, FIGURES WILL BE IGNORED)}
\def\figin#1{\vskip2in}
\else\message{(FIGURES WILL BE INCLUDED)}\def\figin#1{#1}
\fi
\def\Fig#1{Fig.~\the\figno\xdef#1{Fig.~\the\figno}\global\advance\figno
 by1}
%
%
%
%
\def\Ifig#1#2#3#4{
\goodbreak\midinsert
\figin{\centerline{
\includegraphics[width=#4truein]{#3}}}
\narrower\narrower\noindent{\footnotefont
{\bf #1:}  #2\par}
\endinsert
}

\font\ticp=cmcsc10

\def \purge#1 {\textcolor{magenta}{#1}}
\def \new#1 {\textcolor{blue}{#1}}
\def\comment#1{}

\def\\{\cr}
\def\text#1{{\rm #1}}
\def\frac#1#2{{#1\over#2}}

\def\calo{{\cal O}}

\def\roughly#1{\mathrel{\raise.3ex\hbox{$#1$\kern-.75em\lower1ex\hbox{$\sim$}}}}
\font\bbbi=msbm10 
\def\mathbb#1{\hbox{\bbbi #1}}

\def\mthsu{\mathsurround=0pt  }
\def\leftrightarrowfill{$\mthsu \mathord\leftarrow\mkern-6mu\cleaders
  \hbox{$\mkern-2mu \mathord- \mkern-2mu$}\hfill
  \mkern-6mu\mathord\rightarrow$}
\def\overleftrightarrow#1{\vbox{\ialign{##\crcr\leftrightarrowfill\crcr\noalign{\kern-1pt\nointerlineskip}$\hfil\displaystyle{#1}\hfil$\crcr}}}
\overfullrule=0pt

%
%
\lref\AMPS{
  A.~Almheiri, D.~Marolf, J.~Polchinski and J.~Sully,
  ``Black Holes: Complementarity or Firewalls?,''
  JHEP {\bf 1302}, 062 (2013).
  [arXiv:1207.3123 [hep-th]].
}
\lref\Mathurrev{
  S.~D.~Mathur,
  ``Fuzzballs and the information paradox: A Summary and conjectures,''
[arXiv:0810.4525 [hep-th]].
}
\lref\BHQIUE{
  S.~B.~Giddings,
  ``Black holes, quantum information, and unitary evolution,''
  Phys.\ Rev.\ D {\bf 85}, 124063 (2012).
[arXiv:1201.1037 [hep-th]].
}
\lref\SGmodels{
  S.~B.~Giddings,
   ``Models for unitary black hole disintegration,''  Phys.\ Rev.\ D {\bf 85}, 044038 (2012)
[arXiv:1108.2015 [hep-th]].
}
\lref\NLvC{
  S.~B.~Giddings,
  ``Nonlocality versus complementarity: A Conservative approach to the information problem,''
Class.\ Quant.\ Grav.\  {\bf 28}, 025002 (2011).
[arXiv:0911.3395 [hep-th]].
}
\lref\NVNL{
  S.~B.~Giddings,
  ``Nonviolent nonlocality,''
  Phys.\ Rev.\ D {\bf 88},  064023 (2013).
[arXiv:1211.7070 [hep-th]].
}
\lref\AMPSS{
  A.~Almheiri, D.~Marolf, J.~Polchinski, D.~Stanford and J.~Sully,
  ``An Apologia for Firewalls,''
JHEP {\bf 1309}, 018 (2013).
[arXiv:1304.6483 [hep-th]].
}
\lref\HaPr{
  P.~Hayden, J.~Preskill,
  ``Black holes as mirrors: Quantum information in random subsystems,''
JHEP {\bf 0709}, 120 (2007).
[arXiv:0708.4025 [hep-th]].
}
\lref\BHobs{
  S.~B.~Giddings,
  ``Possible observational windows for quantum effects from black holes,''
Phys.\ Rev.\ D {\bf 90}, no. 12, 124033 (2014).
[arXiv:1406.7001 [hep-th]].
}
\lref\BCP{
  A.~Buonanno, G.~B.~Cook and F.~Pretorius,
 ``Inspiral, merger and ring-down of equal-mass black-hole binaries,''
Phys.\ Rev.\ D {\bf 75}, 124018 (2007).
[gr-qc/0610122].
}
\lref\GiShone{
  S.~B.~Giddings and Y.~Shi,
  ``Quantum information transfer and models for black hole mechanics,''
Phys.\ Rev.\ D {\bf 87}, 064031 (2013).
[arXiv:1205.4732 [hep-th]].
}
\lref\Susstrans{
  L.~Susskind,
  ``The Transfer of Entanglement: The Case for Firewalls,''
[arXiv:1210.2098 [hep-th]].
}
\lref\Trieste{
  S.~B.~Giddings,
  ``Quantum mechanics of black holes,''
[hep-th/9412138].
}
\lref\BHIUE{
  S.~B.~Giddings,
 ``Black hole information, unitarity, and nonlocality,''
Phys.\ Rev.\ D {\bf 74}, 106005 (2006).
[hep-th/0605196].
}
\lref\GiShtwo{
  S.~B.~Giddings and Y.~Shi,
  ``Effective field theory models for nonviolent information transfer from black holes,''
[arXiv:1310.5700 [hep-th]], Phys.\ Rev.\ D (in press).
}
\lref\NVNLT{
  S.~B.~Giddings,
  ``Modulated Hawking radiation and a nonviolent channel for information release,''
[arXiv:1401.5804 [hep-th]].
}
\lref\NLEFTone{
  S.~B.~Giddings,
  ``Nonviolent information transfer from black holes: a field theory parameterization,''
Phys.\ Rev.\ D {\bf 88}, 024018 (2013).
[arXiv:1302.2613 [hep-th]].
}
\lref\GWrev{
  J.~Centrella, J.~G.~Baker, B.~J.~Kelly and J.~R.~van Meter,
  ``Black-hole binaries, gravitational waves, and numerical relativity,''
Rev.\ Mod.\ Phys.\  {\bf 82}, 3069 (2010).
[arXiv:1010.5260 [gr-qc]].
}
\lref\BuSa{
  A.~Buonanno and B.~S.~Sathyaprakash,
 ``Sources of Gravitational Waves: Theory and Observations,''
[arXiv:1410.7832 [gr-qc]].
}
\lref\LIGO{
  B.~P.~Abbott {\it et al.} [LIGO Scientific and Virgo Collaborations],
  ``Observation of Gravitational Waves from a Binary Black Hole Merger,''
Phys.\ Rev.\ Lett.\  {\bf 116}, no. 6, 061102 (2016).
[arXiv:1602.03837 [gr-qc]].
}
\lref\LIGOweb{LIGO Open Science Center, ``Data release for event GW150914," \hfil \break https://losc.ligo.org/events/GW150914/\ .}
\lref\GRtests{
  B.~P.~Abbott {\it et al.} [LIGO Scientific and Virgo Collaborations],
  ``Tests of general relativity with GW150914,''
[arXiv:1602.03841 [gr-qc]].
}
\lref\Lehneretal{
  C.~Palenzuela, S.~L.~Liebling, D.~Neilsen, L.~Lehner, O.~L.~Caballero, E.~O'Connor and M.~Anderson,
  ``Effects of the microphysical Equation of State in the mergers of magnetized Neutron Stars With Neutrino Cooling,''
Phys.\ Rev.\ D {\bf 92}, no. 4, 044045 (2015).
[arXiv:1505.01607 [gr-qc]].
}
\lref\pret{
  N.~Yunes, K.~Yagi and F.~Pretorius,
 ``Theoretical Physics Implications of the Binary Black-Hole Merger GW150914,''
[arXiv:1603.08955 [gr-qc]].
}

\Title{
\vbox{\baselineskip12pt  
}}
{\vbox{\centerline{Gravitational wave tests of quantum modifications } \centerline{to black hole structure -- with post-GW150914 update}
}}

\centerline{{\ticp 
Steven B. Giddings\footnote{$^\ast$}{Email address: giddings@physics.ucsb.edu}
} }
\centerline{\sl Department of Physics}
\centerline{\sl University of California}
\centerline{\sl Santa Barbara, CA 93106}
\vskip.10in
\centerline{\bf Abstract}
A  preliminary discussion is given of the prospects that gravitational-wave observations of binary inspiral of black holes could reveal or constrain quantum modifications to black hole dynamics, such as are required to preserve postulates of quantum mechanics.  Different proposals for such modifications are characterized by different scales, and the size of these scales relative to those probed by observation of inspiral signals is important in determining the feasibility of finding experimental signatures.  Certain scenarios with strong quantum modifications in a region extending well outside the horizon are expected to modify classical evolution, and distort the near-peak gravitational wave signal, suggesting a search for departures from waveforms predicted by general relativity.  The near agreement of the GW150914 signal with such waveforms is discussed, and indicates constraints on some such scenarios.  Important strategies for more precise future tests are 1) to develop more precise predictions from scenarios proposing quantum modifications, and 2) searching for observed deviations from numerical relativity predictions via analysis of gravity wave data, particularly focussing on the signal region corresponding to plunge and merger.

\vskip.3in
\Date{}

\newsec{Quantum structure and gravity wave observations}

The unitarity crisis (sometimes called information paradox) for black holes brings into stark focus the need to modify one or more of the cherished principles underlying local quantum field theory.  In particular, no formulation has been found of a local quantum description of black holes, which ultimately respects the principles of quantum mechanics.  

Specifically, a description of black hole (BH) evolution that obeys the unitarity principle of quantum mechanics requires that quantum information that falls into a black hole, or information that is lost to the interior due to the Hawking process, must ultimately escape the black hole.\foot{This statement can be made more sharply in terms of the a required transfer of entanglement from the black hole\refs{\HaPr\GiShone-\Susstrans}.}  This contradicts a local and semiclassical picture of evolution, in which escape of information, which would require its faster-than-light propagation, is forbidden by the locality principle of quantum field theory.  

In order to save quantum mechanics, the picture of local quantum propagation on a  semiclassical spacetime must therefore be modified so that information can escape a black hole, or even more radical assumptions need to be made.  These modifications must extend over a scale at least the size $R$ of the horizon radius, in order to allow transfer of information out of the black hole.  If one takes the viewpoint that we must accept such modifications, but would like to maintain as many of the predictions of semiclassical physics and local quantum field theory as possible, stringent constraints are placed on possible scenarios.  There are also good reasons -- such as the desire to preserve black hole thermodynamics -- to believe that such modifications should in particular take the form of modifications of the semiclassical spacetime description.  

The scales on which such modifications become relevant are clearly important.  One hypothesis\refs{\AMPS} is that they extend over a radial distance $R$ -- from the center of the black hole to the horizon -- but then sharply end there, within a Planck length of the horizon.  This scenario is highly-tuned and appears unnatural, and in fact has been argued to imply a drastic breakdown of spacetime at the horizon\refs{\Trieste,\AMPS}, now commonly called a ``firewall."    Another hypothesis\refs{\BHIUE\NLvC\SGmodels\BHQIUE\NVNL\NLEFTone\GiShtwo-\NVNLT} is that these new effects extend over a range of size $\sim R$ outside the horizon, avoiding such tuning and the violent breakdown of spacetime at the would-be horizon.  

Our focus  is  on whether such scenarios can be observationally probed through careful study of gravitational-wave signals from binary inspiral of black holes.  A discussion of this begins  by characterizing the relevant scales\refs{\BHobs}.\foot{For related comments see \AMPSS.}  One might not expect new effects modifying the semiclassical evolution of a black hole to extend far outside it's strong gravity range; we will call the distance scale outside the horizon on which there are such sizeable ({\it e.g.} $\calo(1)$) modifications to the standard evolution $R_a$, and we thus expect that $R_a$ is of size $R_a\sim R$ or smaller.  A second scale is that on which new effects vary; this is a ``hardness" scale, which we will call 
$L$.  It is certainly possible that $L\sim R_a$ but also one can consider much more rapid variations, $L\ll R_a$.  

To illustrate use of these parameters, the firewall scenario assumes that $R_a\sim L\sim l_{\rm Planck}$, or is determined by some other microscopic length scale.  A less violent hypothesis\refs{\BHIUE\NLvC\SGmodels\BHQIUE\NVNL\NLEFTone\GiShtwo-\NVNLT} is that $R_a\sim R^p$,  $L\sim R^q$, so that the scales increase for larger black holes; the simplest case is obviously $p=q=1$.  Finally, a third kind of proposal has been conjectured, that of fuzzballs\refs{\Mathurrev}, where stringy higher-dimensional geometry becomes relevant outside the would-be horizon.  While no specific description of fuzzball states for Schwarzschild or Kerr geometries has been given,  one expects such a description would have a ``harder" (higher-momentum) character, {\it e.g.} with large average squared curvatures, since these states would involve the higher dimensional geometry and/or stringy structure.  Thus, this third scenario at least raises the possibility that $L\ll R_a$ -- {\it i.e.} $L$ is given by a short distance cutoff scale, but the region over which departures from standard geometry extend ranges much further, even perhaps to an $R_a\sim R$.  

Our ability to test such possible modifications of the geometrical description through gravitational-wave emission depends in part on comparison of these scales with the scales relevant for the binary inspiral.  For simplicity, consider an equal mass binary, with two Schwarzschild black holes each with $R=2m$, which coalesce to form a black hole with radius $R_f \approx 2R$.  Gravitational waveforms for such a process have been well-studied (for reviews see \refs{\GWrev,\BuSa}), but it is important to identify the corresponding configuration of the black holes at different stages in the inspiral.  A useful reference drawing a correspondence between the emitted waveform and this configuration -- based on numerical simulation -- is \BCP.  In outline, the inspiral terminates with a final plunge, once the BHs reach an approximate analog of the innermost stable circular orbit, at separation $d\sim 12R$.  Here they then plunge towards merger, via formation of an apparent horizon, expected to happen at $d\sim 4R$.\foot{Care is needed to make a gauge-independent statement\BCP.} These happen a time $\roughly> 12 R$ and $6R$, respectively, before the peak power output of the gravitational-wave signal, in rough consistency with the observation that even at apparent-horizon formation, the tangential velocity still dominates radial velocity.  

Comparing the scales of proposed quantum modifications to geometry to those of inspiral gives us important information about whether one can look for signals to observe or constrain different quantum scenarios.  For example, if all modifications are confined to a region within $R_a\sim l_{\rm Planck}$ of the horizons of the infalling black holes, as for firewalls\AMPS,  these modifications should not reach out to influence the inspiral/plunge;  and once an apparent horizon forms one expects it to encompass the original horizons, cloaking any possible signal from the new structure.  

The more natural case $L\sim R_a\sim R$\refs{\BHIUE\NLvC\SGmodels\BHQIUE\NVNL\NLEFTone\GiShtwo-\NVNLT} clearly offers more promise for observation/constraints.  If, for example, $R_a=R$, extending just past the light-orbits of the infalling black holes (which are at Schwarzschild coordinate $r=3R/2$), and if there are significant modifications to the geometrical description here, these modifications are expected to influence the mutual motion of the black holes, and the gravitational-wave signal in the final part of the plunge to near apparent-horizon formation at $d\sim 4R$.  

Alternately, in the case of macroscopic, hard structure ($R_a\sim R$, $L\ll R_a$), which might be conjectured to arise from fuzzballs, one expects interaction of this structure should lead to hard/high-energy components of the radiation from the merger.  

Already, then, first signals from binary mergers should start to indicate tests for quantum modifications needed to save unitarity.  For example, certainly an unexpected component of hard radiation near merger (if resolvable) would point to hard structure, and its absence conversely would constrain such structure.

But what about the case of soft, but strong, modifications to gravity, that have been proposed in \refs{\BHIUE\NLvC\SGmodels\BHQIUE\NVNL\NLEFTone\GiShtwo-\NVNLT}?  Here, if strong modifications to the geometry extend to sufficient distances $R_a\sim R$, one expects one could see modifications to gravitational wave signals near the peak in energy, from the region where the BHs approach merger and the quantum-modified regions overlap.  Specifically, for sufficiently strong modifications one would expect significant departures from the classical general relativity (GR) predictions, in this region of the signal.

Another potential window of opportunity is offered by the ringdown of the final black hole, governed by the quasinormal modes in the standard scenario.  If there are new quantum effects, these could produce anomalies in the ringdown dynamics.  However, there is a bigger uncertainty\BHobs\ -- that of when such new effects manifest themselves for a newly-formed black hole.  If new effects outside the horizon take a time that is longer than the ringdown to reach significant size, they would not then be visible.  In contrast, while there are similar timescale uncertainties with the initial black holes, these will have existed for a much longer time, decreasing this kind of uncertainty.

Obviously more quantitative statements are strongly desired.  One approach to describing, in an effective field theory 
approach\refs{\NLEFTone\GiShtwo-\NVNLT}, the possible strong-but-soft modifications to classical geometry is through deviations from the classical metric,
\eqn\effmet{g^{\mu\nu}= g_{\rm class}^{\mu\nu} + H^{\mu\nu}\ ,}
as discussed in \BHobs. One can think of these as fluctuations in the effective metric that depend on the quantum state of the black hole.
One wants to consider perturbations that are nonsingular on the would-be horizon, so a useful parameterization uses coordinates like the Kerr coordinates $(v,r,\theta,\tilde \phi)$ (where $v$ is an advanced coordinate) which are regular there, {\it e.g.} taking the general form
\eqn\kpert{H^{\mu\nu} \sim f(r,v) e^{-i\omega v+ ik r} Y_{lm}(\theta,\tilde\phi)\ .}
Here $f$ is a window function which localizes to the region of extent $R_a$ near the horizon; $k\sim 1/L$ models the ``hardness" of the perturbation.  
An approach to characterizing the sensitivity of the gravitational-wave signal to such new effects is to initialize simulations with data where the incoming black holes, near merger, have such perturbed metrics.  Of course, we are still lacking a more fundamental framework to describe the full dynamical evolution responsible for such perturbations.  But, an initial understanding of sensitivity to their effects would be expected to be found by considering such data, enforcing the constraints, and following its standard Einsteinian evolution.  For example, in the case of test probes (massive or massless) of a single black hole, ref.~\BHobs\ argued that such perturbations of sufficient strength lead to $\calo(1)$ deviations in the geodesics over a range comparable to an orbit.  While analyzing the analogous detailed effects in the strong region of merging black holes is clearly a formidable numerical task, one expects similar statements to hold there.  

Then, one expects possible $\calo(1)$ deviations from the classical geometry of merger, with commensurate contribution to the gravitational wave signal.  In particular, the fact that such perturbations generically decrease the symmetry of the problem strongly suggests a more irregular signal and increased power in the near-peak signal.  This points to the importance of a search for such anomalies, which could give the first indication of a discovery of quantum black hole structure, and provide strong impetus to a more careful analysis.

\newsec{Post-GW150914 update}

The preceding discussion was written prior to LIGO's  announcement of the GW150914 detection\refs{\LIGO}, and so it is important to assess the situation in light of this historic observation.  In short, the GW150914 event provided data that was very close to numerical predictions of classical general relativity.  One way to see this is to look at the residuals in fig.~1 of \LIGO.  To suppress noise from individual detectors, these residuals are multiplied, using the publicly-available raw data\refs{\LIGOweb}; the result is shown in fig.~1.  

\Ifig{\Fig\figlabel}{The product of the residuals for GW150914 from the Hanford and Livingston detectors is shown in black; also shown, for reference, are the individual Hanford and Livingston signals. The  region corresponding to plunge and merger (generously, $0.37-0.43\,s$ in the plot) is indicated.  The residual here is similar in character to that in the earlier inspiral region; compare, {\it e.g.}, the range up to $0.35\, s$.  This indicates consistency of the residual with noise, rather than it being due to signal effects due to GR modifications during plunge and merger, at the level of sensitivity accessible via this signal.  (Plot by S. Koren.) }{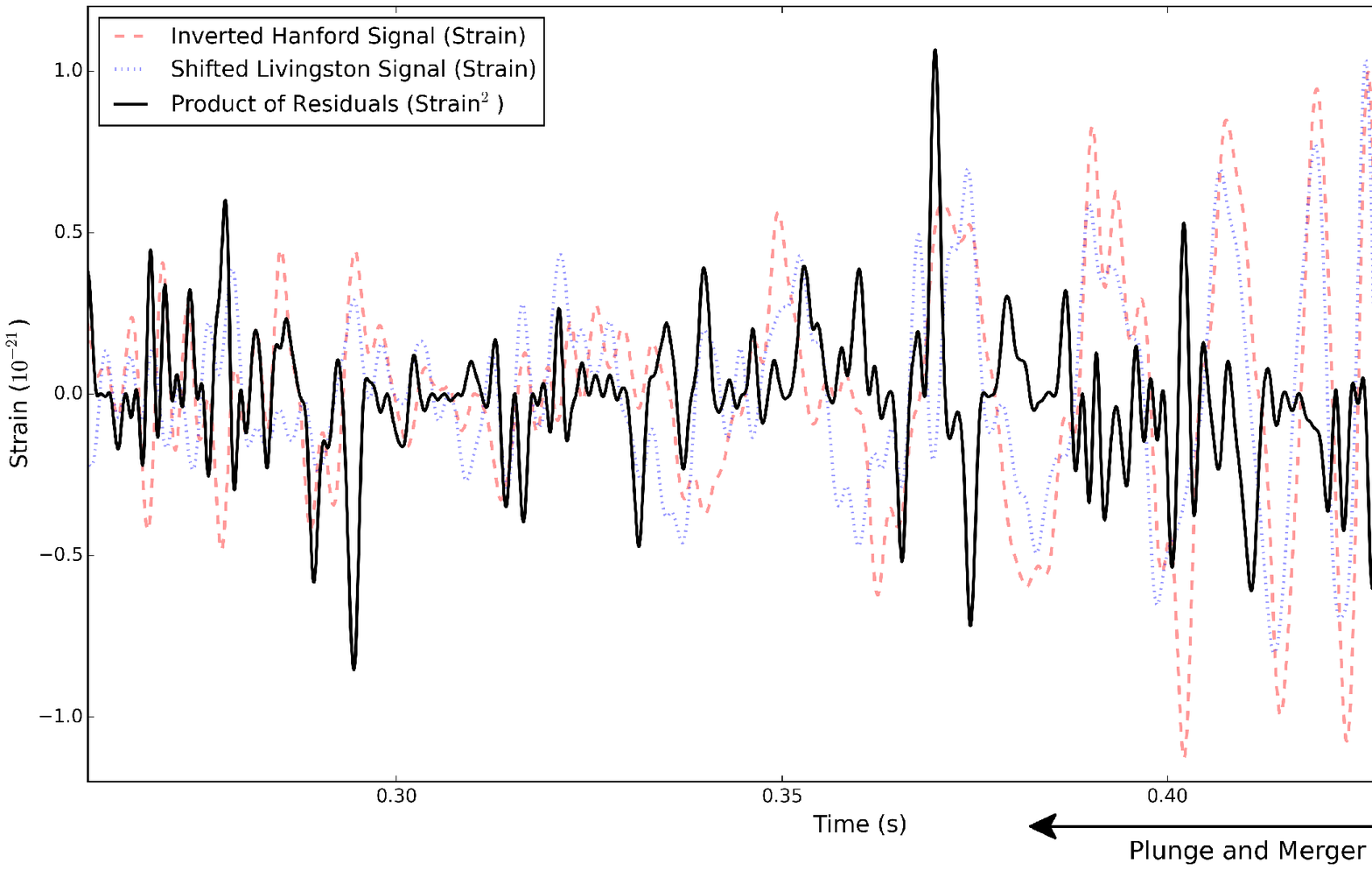}{7}

An important point is that the combined residuals (black curve) in the vicinity of plunge/merger are very similar to those in the earlier inspiral regime, as can be seen from direct inspection of the figure.  This suggests compatibility of the observed residual with noise, as opposed to a significant signal deviation from classical GR.  
For example, the largest spike in the combined residuals falls outside the plunge/merger regime.  The spikes in the plunge/merger regime have amplitudes, widths, and spacings that are similar to those in the pre-plunge regime, as can be seen by eye.  Given the limits arising from the use of one event, it may be difficult to go much further with a more sophisticated analysis, but once more events are found where there is sensitivity to these phases, it may be possible to apply more sophisticated statistical tests for small departures.
Other tests for departures from GR include comparison of the final BH mass to that predicted by GR, given the measured inspiral parameters, with good agreement shown in \refs{\GRtests}.  Thus, clearly this first gravity wave detection does not reveal large departures from classical GR.

An example of the possible effects of departures from classical vacuum BH  mergers is provided by the case of neutron star(NS) mergers; these illustrate some kinds of deviations that can occur.  A recent nice discussion of the effects of different NS equations of state on gravitational waveforms is given in \refs{\Lehneretal}.  The neutron star radii are analogous to the radius $\sim R+R_a$ described in the preceding section.  Once the neutron star surfaces make contact, non-gravitational scattering begins and the waveform quickly departs from that for black holes.  This is most clearly seen in fig.~2 of \Lehneretal; there, the least compact NS departs first from the BH curve, followed by the solutions that correspond to more compact NSs.  The departures are clearly significant, {\it i.e.} $\calo(1)$.  

Let us compare the NS example to BH quantum structure.  First, a NS surface is significantly larger than $R$; for the equations of state considered in \Lehneretal\ the range of radii is $2.4R-3.2R$.  Quantum modifications to classical spacetime may certainly, as described above, have  a more limited range.  For an extreme example, the ``firewall\AMPS" case $L\sim R_a\ll R$ certainly appears to be consistent with the LIGO signal, as the regions of deviation  do not meet until the BHs essentially ``touch," and so can remain inside the final BH.  

Structure with scales $L\ll R_a\sim R$, on the other hand, would be much more ``neutron-star-like," in that  the ``hard" structures (rapidly varying in space, on scale $L$) of the two objects collide and are generically expected to lead to additional scattering.  Note that while such scattering is expected to produce significant waveform distortion, as with the NS case, this distortion can depend on the exact description of the quantum BH structure.  In the NS case, initially the radiated power is reduced, as the energy of collapse is diverted into other channels ({\it e.g.} distorting the individual NSs), reducing the gravitational signal.  However, 
if the quantum structure in the BH case is intrinsically gravitational ({\it e.g.} rapid variations in the gravitational field), that suggests additional channels for gravitational radiation, increasing the signal.  Even in the NS case, such predictions depend on models of the NS structure, so, in short, precise predictions depend on a more precise description of the quantum structure and its evolution.

One should note that LIGO sensitivity to such structure may also be reduced by two effects.  The first is that if the quantum BH structure is more localized than that of a NS, its scattering creates perturbations that originate closer to the horizon of the final BH.  Such perturbations are more readily absorbed by the final BH, rather than escaping to infinity, decreasing their effect on the signal.  Secondly, the GW1509014 signal is bandlimited, largely due to noise considerations; for example the data in fig.~1 of \LIGO\ is filtered to $<350$Hz.   Thus, sensitivity is lost to higher-frequency components, if present.  

Similar considerations apply to the case $L\sim R_a\sim R$.  Here one also expects deviations in the gravitational-wave signal, but that are smaller and lower-frequency due to the ``softer" structure.  In addition, in the case where $R_a\roughly<R$, much of a signal may be absorbed into the final BH, in line with the preceding comments, reducing observability/power to constrain.

Nonetheless, it appears that GW150914 {\it has already} provided constraints on scenarios with sufficiently long range and hard structure, {\it e.g.} comparable to that of neutron stars (rescaled to the relevant mass range, $M\sim 30 M_\odot$).  To see this, note that if one rescaled all dimensionful parameters to replace the $\sim1 M_\odot$ NSs of \Lehneretal\ with  $M\sim 30 M_\odot$ objects, the departures shown in fig.~2 of \Lehneretal\ would be evident in fig.~1 of the present paper, since the departures of the former are large in the pre-merger phase.  This of course corresponds to using a fictitious equation of state, that allows $\sim 30 M_\odot$ analogues of NSs.  But, this in turn suggests one approach to beginning to parameterize departures from GR, and analyze their possible effects.  One can consider model equations of state that replace BHs with exotic objects with structure outside the would-be horizon.  
Then one can analyze their effects on the GW signal, in analyses extending \Lehneretal\ and other similar work.

For example, while no quantitative predictions are yet made for Schwarzschild/Kerr black holes in the conjectured fuzzball scenario\Mathurrev, one generically expects microstructure at $L\ll R$ for these; such microstructure of sufficient magnitude, if it extended to $R_a\sim R$, is apparently constrained by the close agreement of 
GW150914 with classical GR -- indicating absence of scattering of hard structure -- unless even further conjectures ({\it e.g.} fuzzball complementarity\Mathurrev) are added.  One possible approach to making this statement more quantitative could be to find an effective equation of state that approximated such fuzzball dynamics.

In conclusion, this paper has argued that the parameters $L,R_a$ are important characteristics of possible quantum structure of black holes, and should be a target for more precise descriptions of such structure.  Of course, ultimately quantitative constraints appear to rely on understanding the dynamical evolution of quantum structure that would replace classical gravity in the strongly nonlinear regime.  This evolution, and study of its departure from general relativity, is needed to actually compute waveforms that could be compared with data.  Thus, a first part of a strategy to investigate possible quantum black hole structure is to more precisely characterize its possible nature.  An intermediate, which may have some phenomenological utility in semiquantitative description of sensitivity of gravitational wave data, may be parameterizing deviations from classical geometries, as indicated in the preceding section, and following their classical evolution.  Alternatively, as is outlined above, one can parameterize certain departures from usual vacuum GR in terms of an effective stress tensor and equation of state, as is done for neutron stars, and this is a practical approach to begin to investigate sensitivity of GW signals to structure that replaces a BH horizon.

A second important element of a strategy for investigating possible quantum black hole structure is, however, clearly evident -- with the expectation of more gravitational wave observations, all reasonable effort should be made to perform analysis that would reveal departures from general relativity predictions, particularly due to modifications in the plunge/merger phases.  While some constraints on this evolution have been noted, much of the work on constraining departures from GR has focussed on departures over the entire inspiral\GRtests, {\it e.g.} due to modification of PPN parameters. (However, see also \refs{\pret} for some more general discussion, in particular noting the need of any new dynamics to mimic the high effective viscosity of GR).  It is particularly important to seek  sensitive tests of  the strong gravity regime, and this indicates a need to focus on constraining or observing any  small deviations from the signal predictions of general relativity that become manifest {\it during the corresponding plunge and merger phases.}

\bigskip\bigskip\centerline{{\bf Acknowledgments}}\nobreak

I thank F. Pretorius for providing background on numerical simulations, L. Lehner for helpful conversations on neutron star waveforms, and D. Holz, G. Horowitz, D. Marolf, and M. Srednicki for useful conversations.  I also thank S. Koren for producing the plot of fig.~1.
This work  was supported in part by the Department of Energy under Contract No. DE-SC0011702, and by Foundational Questions Institute (fqxi.org) 
grant FQXi-RFP-1507.

\listrefs
\end